\documentstyle[preprint,aps]{revtex}
 
\begin{document}

%\draft

%\def\thefootnote{\fnsymbol{footnote}}

\preprint{hep-th/9611072,  
EFI-96-41,
UT-KOMABA/96-22}
 
\title{
D-Particle Dynamics and The Space-Time Uncertainty  
Relation
}
 
\author
{Miao Li
\footnote{
{\tt mli@curie.uchicago.edu}}
}

\address
{Enrico Fermi Institute, University of Chicago\\
5640 Ellis Avenue, Chicago, IL 60637
}

\author{Tamiaki Yoneya
\footnote{
{\tt tam@hep1.c.u-tokyo.ac.jp}}
}
\address{Institute of Physics, University of Tokyo\\
Komaba, Meguro-ku, 153 Tokyo
}

\maketitle
\begin{abstract}
We argue that the space-time uncertainty relation of the
form $\Delta X \Delta T \gtrsim \alpha' $ for the observability of
the distances with respect to time, $\Delta T$, and space, $\Delta  
X$,
is universally valid in string theory including D-branes.
This relation has been previously
proposed by one (T.Y.) of the present authors as a
simple qualitative representation of the
perturbative short distance structure of
fundamental string theory. We show that the relation,  
combined with the
usual quantum mechanical uncertainty principle,
explains the key qualitative features
of D-particle dynamics.
\end{abstract}


\newpage
It is often stated that in the fundamental string theory
there exists a minimum length of order of $\sqrt{\alpha'} \equiv  
\ell_s$
beyond which we cannot probe the structure of space-time and hence
the ordinary concept of space-time ceases to be meaningful.
This comes about from the properties of string
amplitudes in the high-energy limit  
\cite{amaticiaven}\cite{grossmende}
and also in the high temperature limit \cite{attickwitt}.
Such a statement is indeed quite natural when we have only
the ordinary string states as possible probes for
short distances, since string states themselves have an intrinsic  
extension of
the order of length $\ell_s$.

Recently, however,  we understood that string theory in fact allows  
a
variety of objects of various dimensions as solitonic excitations  
and that
they are bound to play crucial roles in nonperturbative
formulations of string theory.  In particular, we have even
point-like objects called D0-branes \cite{pol1} or D-particles.
Recent studies \cite{barbon}\cite{bachas}\cite{lifsh}
\cite{daniel1}\cite{kaba}\cite{dkps}\cite{daniel2}\cite{lifsh2}
 of D-particle dynamics revealed the
possibility of probing the distance scales of 11D Planck scale
of the order $g_s^{1/3}\ell_s$, the  natural scale of the
M-theory \cite{witt}, which is indeed
much shorter than the string scale $\ell_s$ for weak string  
coupling
\footnote{
The importance of shorter length scales in string theory has been
suggested earlier in \cite{shenker}. }.
Therefore, the usual folklore statement quoted above must now
be reconsidered.  If we remember that
the string scale represents the unique fundamental constant of  
Nature
in the natural unit $c=\hbar=1$,
its precise significance must certainly be clarified.
The purpose of the present note is to remark that a simple  
space-time uncertainty
relation \cite{ty1} proposed in 1987
by one of the present authors
 is an appropriate interpretation for the meaning of the string  
scale $\ell_s$,
since it is universally valid both for ordinary string scattering  
and
D-particle dynamics.

Let us begin with briefly recalling the arguments of reference  
\cite{ty1}
which motivates the space-time uncertainty relation.
Consider a high energy scattering of arbitrary objects whose  
interactions are
mediated by strings. If the energy scale is of order $E$, the
smallest time scale probed by this scattering event is of order
$\Delta T \sim {1\over  E}$.  Now, what is the typical
spatial length scale probed by this scattering event?  If both the  
scattering
objects and their
interactions were described by usual local field theories
neglecting  quantum gravity, we would be
allowed to state that it is determined by the typical wave length
of the objects, namely,  ${1\over E}$ for sufficiently high  
energies.
Hence, in principle, we  would
have no limitation for probing the short-distance scale,
provided we neglect quantum gravity. If on the other hand the  
interactions
are mediated by fundamental strings, high energies  do not
necessarily imply that the typical spatial scale is given by the
wave length of the scattering objects, since higher energies
dominantly cause larger fluctuations with respect to string  
excitations
during interactions than with respect to the center of mass motion  
because of
the huge degeneracy of string excitation modes.
It is easy to see \cite{ty1}\cite{ty2}
that the typical (smeared-out) spatial extension $\Delta X$
of strings with energy $E$ is of order $\Delta X \sim \ell_s^2E$.
This implies the simple relation for the
indeterminacies of the space and time lengths
\begin{equation}
 \Delta X \Delta T\gtrsim \ell_s^2, 
\label{st-relation}
\end{equation}
which we call the space-time uncertainty relation.
In reference \cite{ty1}, this relation was proposed as a natural  
space-time
representation of the $st$-duality properties of string scattering  
amplitudes.
As  discussed later in reference\cite{ty2}, it can also be  
derived as
a direct consequence of the world-sheet conformal invariance.

      From the view point of the space-time uncertainty relation,
the usual argument \cite{review} for the
minimal length essentially
amounts to assuming that the observable length is the average
of the spatial and time distances; then
we would have  the lower bound ${\Delta X + \Delta T \over 2} \ge  
\ell_s$.
Clearly, however, what is the dominant scale measured by scattering
experiments depends on which kinematical regions we are interested  
in.
For example, a high-energy low-momentum transfer (peripheral)  
scattering
experiment can probe small time scale, but the spatial scale
is not necessarily small.  Thus, it corresponds to $\Delta  
T\rightarrow 0$ and
our relation (\ref{st-relation}) implies that the spatial length scale  
grows as
${\ell_s^2 \over \Delta T}\sim \ell_s^2 E$ as the laboratory
energy $E$ increases.  By adapting the string-bit argument due to
Susskind \cite{sk}, this longitudinal
length scale of a string, 
growing linearly with energy, leads to the Regge  
intercept $\alpha(0)=2$ of
the probability amplitude $A(E) \sim
E^{\alpha(t)-1}$, in conformity with the
existence of a graviton
\footnote{
Here, $t$ is the invariant momentum transfer whose dependence
is basically determined by the effective transverse size of the  
string.}
.  On the other hand, a high-energy fixed-angle
scattering as studied in \cite{grossmende} tries to probe short  
distances
with respect to both space and time.  In this case,  since the  
relation
(\ref{st-relation}) shows there is no such degrees of freedom, the
amplitude vanishes exponentially in the high-energy limit.
 
As long as we only use the strings as probes, it seems difficult to
imagine a scattering experiment which makes it possible to measure
directly the
region $\Delta X\rightarrow 0$, because of the intrinsic extension
of the string. In references \cite{ty1}\cite{ty2}, it was suggested  
to
interpret the relation (\ref{st-relation}) in the limit $\Delta  
X\rightarrow 0$
as an explanation why it is possible to treat the asymptotic
string states, propagating
infinitely long time $\Delta T \rightarrow \infty$,
as {\it local} external fields  in the sigma-model approach to
world-sheet string theory.  The asymptotic states correspond
to the $s$-channel poles. Combined with the Regge-pole 
exchange picture for the limit $\Delta T \sim {1\over \Delta X}  
\rightarrow 0$,
the relation (\ref{st-relation}) is thus interpreted as a natural  
space-time
interpretation of  [Regge-pole]$\leftrightarrow$[resonance-pole] duality.
Now it is clear that the D0-branes are
ideal objects for the purpose of directly probing the  short  spatial length scale
and testing the relation (\ref{st-relation})
beyond such formal arguments.
Fortunately, there already appeared several works cited above which  
studied
the dynamics of D0-branes in the low-energy limit.
In the following, we show that all the results so far
are consistent with the relation (\ref{st-relation}) and
the most crucial feature behind these results can be
naturally understood on the basis of (\ref{st-relation}).

First of all we note that
the slow velocity limit studied in these works is
just appropriate for  probing the
small $\Delta X$ regions where (\ref{st-relation}) implies
that $\Delta T$ grows.  Moreover, it seems fairly clear that the
space-time uncertainty relation just conforms to
the basic principles emphasized in reference \cite{dkps}
that the leading singular behavior in the short spatial distance  
limit
is determined by the {\it IR} behavior of brane world-volume quantum  
theory.
In fact this statement
 is a direct consequence of the duality between $s$-and-$t$ channels
 which was nothing but
the original interpretation of the relation  (\ref{st-relation}) as  
discussed above.
In this sense, our discussion  will provide a shortcut to  
understanding
some of the basics for previous results on the short distance  
behavior
of D-particle dynamics.

Now let us consider the scattering of two heavy D-particles of
 mass $m= {1\over g_s\ell_s}$ with slow typical velocity $v$.
If we assume that there is a limitation for the meaningful  
space-time lengths
in the form (\ref{st-relation}), what is the smallest possible
spatial length scale $b$ probed by this scattering?
Let
$$
b\sim v^{\eta}\ell_s.
$$
The typical time length of the scattering is $t = {b \over v}$.
Substituting these relations to (\ref{st-relation}) with
$\Delta X \sim b, \Delta T \sim t$, we must have $\eta =1/2$,
\begin{equation}
b \sim \sqrt{v}\ell_s.
\label{stadiumsize}
\end{equation}

This length scale first appeared in \cite{bachas}, where
an annulus amplitude of open strings is computed, and
was further analyzed in detail in \cite{dkps} using the
effective world line theories \cite{witt2} 
($0+1$-dimensional super Yang-Mills  
matrix quantum mechanics)
of D-particles.
The relation (\ref{stadiumsize}) constitutes one of the most
crucial relation in all previous discussions of short-distance
structure in D-particle dynamics.  Following \cite{dkps}, we often
call the scale $\sqrt{v}\ell_s$ the stadium size. From the point of view of
the effective super Yang-Mills matrix model, the stadium size is the
limit for the spatial scale where the Born-Oppenheimer
approximation for the coupling between the D-particle coordinates
and the short open-string excitation connecting them ceases to be  
valid.
The dynamics of latter corresponding to the
off-diagonal part of the adjoint Higgs fields originated
from the 10D gauge fields by dimensional reduction
is governed by the time scale
${\cal O}(\ell_s^2/b)$ while that of the former, the diagonal
part of the Higgs, is by ${\cal O}(b/v)$.
If the Born-Oppenheimer approximation is not valid, we cannot
clearly separate the diagonal part as the spatial coordinates of  
D-particles and
hence we must have (\ref{stadiumsize}).  Our proposal is
thus to interpret this result as a consequence of the universal
space-time uncertainty relation (\ref{st-relation}).

Now once the relation (\ref{stadiumsize})
is known, the important fact that the characteristic spatial length  
scale of D-particle dynamics is nothing but the eleven dimensional
Planck scale associated with M-theory is understood from the
usual uncertainty relation.  If the time duration of the scattering  
is of order $\Delta T\sim {b\over v} \sim v^{-1/2}\ell_s$,  
usual time-energy uncertainty relation applied for a point-like
D-particle implies 
an uncertainty with respect to the D-particle 
velocity of order $\Delta v\sim g_sv^{-1/2}$  
which leads to the spread of the wave packet of order 
$g_s v^{-1}\ell_s$ during the time interval $v^{-1/2}\ell_s$.   Here   
we used the fact that the kinetic energy of a 
D-particle is ${1 \over 2g_s\ell_s}v^2$ for weak string coupling
\footnote{
For the present order estimate, we can neglect the potential energy  
of order
${\cal O}({v^4 \over (\Delta X)^7})\ell_s^6$ which is smaller than the 
kinetic energy when $\Delta X \gtrsim b$. 
}.
In oder that the minimum length scale $b\sim \sqrt{v}$
be meaningful, $b$ must be larger than this spread.  Thus we have
the lower bound for the velocity \cite{dkps}
$$
v \gtrsim g_s^{2/3}
$$
which leads to the 11D Planck scale $b\sim g_s^{1/3}\ell_s$ of the
M theory as a meaningful smallest distance 
probed by  D-particle scattering at low velocities.  
     
        From the view point of the effective super Yang-Mills matrix model, 
the 11D Planck scale is easily understood from a scaling argument 
\cite{daniel1}\cite{kaba} which says that coupling constant $g_s$ 
can be eliminated from the dynamics by making  a rescaling,  $X_i \rightarrow g_s^{1/3}X_i$ and $t \rightarrow g_s^{-1/3}t$, 
for the D-particle coordinates 
$X_i$ and the time $t$, respectively.  We here emphasize 
a trivial but crucial fact that the {\it opposite}  
scalings for the space and time coordinates just conform to  
a necessary requirement 
for the validity of  the relation (\ref{st-relation}).

Actually, as pointed out in \cite{dkps}, when we consider
a D-particle in the presence of a large number of
D4-branes,  it becomes possible to probe arbitrary short spatial  
distance scale.
This is basically due to the fact that the D4-branes produce an  
effective metric
for the moduli space of a D0-particle which
makes the effective mass of D0-particle much heavier than that in
the flat space at short distances: The effective action
in the presence of $N$ coincident parallel D4-branes is given by
$$
S_{{\rm eff}}= \int dt [{1\over 2g_s\ell_s}(1+{Ng_s\ell_s^3 \over  
r^3})v^2 +  {\cal O}(Nv^4 \ell_s^6/r^7)]
$$
where $r$ is the distance between the D-particle and D4-branes.
According to \cite{dkps}, this metric is likely to be exact
without $\alpha'$ and instanton corrections.
When the distance $r$ is much shorter than $(Ng_s)^{1/3}\ell_s$,
the mass of the D-particle is effectively given by
$m\sim N\ell_s^2/r^3 \gg {1\over g_s\ell_s}$. We can then easily  
check that  the spread of the D-particle wave packet can be neglected
during the time $t \sim v^{-1/2}\ell_s$ compared with the stadium  
size 
$\sqrt{v}\ell_s$ for large $N$ (if $r$ itself is the stadium size 
$\sqrt{v}\ell_s$). This allows us to probe arbitrary  
short lengths
with respect to the distance between the D-particle and D4-branes  
and
hence indicates that the singular spatial metric in the D0-particle
moduli space is meaningful even in the limit $r\rightarrow 0$.
On the other hand, since the time scale grows indefinitely,
we cannot talk about the interaction time in any meaningful way
in the limit of short spatial distance.

The space-time uncertainty relation (\ref{st-relation})
in general says that to probe the 
short spatial distances, inversely large time interval is necessarily   
required and 
vice versa. 
Thus if it is universally valid, we would not be able to  introduce the   
concept of 
space-time event which is local with respect to both space and time.   
 
Although we expect that the space-time locality is
lost eventually in any theory including quantum
gravity,  the relation (\ref{st-relation}) suggests a
specific manner on how this happens in fundamental string theories.
It is very important to see whether this way of expressing the
significance of the string constant $\ell_s$ is useful
in the dynamics of more general branes and strings,
including D-instantons.  Since the interaction between
general D-branes is governed by the fundamental strings
and the relation (\ref{st-relation})  originates from the
conformal invariance of the fundamental string dynamics,
it is reasonable to expect its general validity, 
 if we interpret the relation appropriately. 
Here, the case of D-instantons (D-1-branes)  is very special
since we cannot talk about time evolution in their dynamics.
It is known \cite{klebthor} that the invariant scattering amplitude
of massless
string states off a fixed D-instanton for arbitrary energy
is simply given by  ${1\over t}$
apart from the kinematical factor
where $t$ is the invariant $t$-channel energy, provided we supply 
appropriate fermion contribution to cancel possible fermion zero modes.
The pole $t=0$ represents the massless dilaton exchange
contribution.  Namely, the amplitude for weak string coupling is  
reproduced
by a local field theory without any $\alpha'$ correction for any
energy.  This is {\it not} a contradiction to our space-time  
uncertainty relation,
since the above behavior can be interpreted to correspond to the  
special case where
$\Delta T \sim {1\over \Delta X} \sim 0$:
$\Delta T\sim 0$ reflects the
point-like nature of the D-instanton in space-time, while
$\Delta X \rightarrow \infty$ is associated with the long range
propagation  of a virtual dilaton exchange.
Indeed, after the integration over the position of a single 
 D-instanton, we would have 
$t=0$.    From the view point of (\ref{st-relation}),
the appearance of the long range exchange of massless dilaton is  
a necessary condition for the existence of the point-like
instanton ($\Delta T\sim 0$) contribution.  Therefore, the
apparent loss of stringy property of the
D-instanton-string interaction may be regarded
as yet another piece of evidence
for the universal nature of the space-time
uncertainty relation. 
Of course, this argument is restricted on one instanton case. 
The multi-instanton dynamics is much more 
complicated  due to stringy interactions and 
the integration over the collective coordinates.  
Further investigations on general D-brane dynamics 
including instantons  from our view point will be 
useful for clarification towards more precise and general 
interpretation of the space-time uncertainty relation.

Our discussions so far assumed weak string coupling.
However, given that, except for type IIA theory and heterotic
$E_8\times E_8$ theory,
the S-duals of  string theories are 
again string theories, it is natural to expect that the 
relation is valid even at strong string coupling. 
Note that the string constant $\ell_s$  can be regarded to be 
 invariant under the 
S-dual transformations, and therefore the 
string tension with respect to the correctly rescaled 
space-time coordinates remain the same as well.  
In the case of type IIA (which we are mainly concerned with in 
the present note) and heterotic $E_8 \times E_8$,     
the dual is the M-theory, whose strong (string)  
coupling limit is believed to be described by 11D supergravity
in the long-distance limit.  If the recent interesting conjecture 
\cite{bfss} that
the microscopic  M-theory is described {\it exactly} by
 the $0+1$D  Yang-Mills matrix model in the infinite-momentum frame  
is correct, it is plausible that our uncertainty relation continues to 
be valid even for strong string coupling, 
since the time scale in the Galilean dynamics 
of the off-diagonal elements 
is then always given by the difference of the diagonal elements 
${\ell_s^2\over |x_i-x_j|}$ for the D-particle coordinates $x_i,  
x_j$,
which is the basis for the relation (\ref{st-relation}) from the 
view point of the super Yang-Mills matrix model as a 
world line theory of D-particles.

Another relevant question towards a possible generalization 
of the space-time uncertainty relation would be whether we  
can have similar relation for spatial domains without 
including the time length. 
The appearance of the lower bound $\Delta X \sim b \gtrsim  
g_s^{1/3}\ell_s$
as the stadium size already suggests that 
the minimum size of  spatial domains is in general 
determined by the 11D Planck length $g_s^{1/3}\ell_s$. 
As a simple but different example, 
let us consider a measurement of  the position 
of a D-particle inside  D4-branes by scattering with an 
external D-particle.   
Classically, a bound state of a D-particle 
with $N$ coincident D4-branes can be described by an instanton 
solution of the D4-brane gauge field \cite{mike}. Probing the D-particle 
inside  these D4-branes is equivalent to probing the localization of the 
background field $A$. The localization of $A$ is disturbed by the 
massless D4-brane open string modes as decay product of pairs of 
open strings connecting the D-particle probe and D4-branes. We note 
that D4-brane open string modes are inevitable product if the  
impact parameter is sufficiently small. The dissipation rate is calculated 
in \cite{bachas} for D-particle-D-particle scattering, and in \cite{dkps} 
for D-particle-D4-brane scattering.  
When the impact parameter is of 
order $\Delta X$, the space-time uncertainty relation shows that 
the typical energy transferred to the massless open string 
modes of the D4-brane is of order $\Delta E \sim \Delta T^{-1} 
\sim {\Delta X\over \ell_s^2}$. If this energy is used for the interaction 
with the D-particle inside the D4-brane to probe its position, it 
would contribute to an uncertainty of velocity of the 
D-particle inside of order $\Delta v ={g_s\ell_s \over v\Delta T} $,   
which implies uncertainty of the position of order 
${g_s\ell_s\over v}$ during the interaction time $\Delta T$. 
On the other hand, the D-particle inside the D4 brane travels the 
distance of the order $v\Delta T$. Thus the net uncertainty of the   
position of the D-particle inside 
the D4-brane is at least of order 
$\Delta Y \sim ({g_s\ell_s\over v}+ v\Delta T)/2 \ge 
\sqrt{\Delta T \ell_s g_s} $ 
\footnote{This argument can be made more accurate 
by using the distribution $\int p^3dp$ for a localized wave function, 
the result will remain the same.}.  
Combined with (\ref{st-relation}), this suggests the validity of a   
strange relation of the following type,  
$\Delta X\Delta Y^2 \gtrsim g_s \ell_s^3$.  
When the D-particle-4-brane impact parameter $\Delta X$ is very small,    
a large lump of massless 4-brane modes is produced and the D-particle 
inside the 4-branes becomes difficult to locate, $\Delta Y \gtrsim  
({g_s\ell_s^3 \over \Delta X})^{1/2} $.  
The 11D Planck scale appeared here again, 
but the argument is apparently 
independent of the  discussion for the minimum stadium size 
for the projectile D-particle.   
We do not know whether this new relation has a universal meaning beyond 
the situation discussed here.  
The origin of this relation might be related with the holographic   
principle
\cite{holographic}.  We should note here 
that, depending on various situations 
such as the dimensionalities of  D-branes
 and the directions of the spatial distances we are interesting in, 
the characteristic 
spatial scale can vary from case to case.  
This is easily seen from the scaling arguments 
of the SYM matrix models.  We want to 
emphasize that, in contrast to this, the 
relation (\ref{st-relation}) for the transverse distances between 
D-branes and the time scale is universal 
since the transverse distances 
and the time always have opposite scaling behavior.  
As is seen from the examples discussed above, we expect 
that most of the space-space relations are derived 
on the basis of the space-time relation, appropriately combined with 
the usual quantum mechanical uncertainty principle. 

     From formal point of view, what is lacking in our arguments is a 
mathematical foundation for the appearance of the uncertainties. 
For example, we cannot  at present 
give any precise quantitative definitions for   
the space and time uncertainties.  We feel that there should be a   
formulation of the
theory which surpasses strings and D-branes. 
 In this respect, a very urgent and challenging problem seems to 
construct a completely covariant formulation 
of interacting D-particle dynamics as an extension of the SYM models  
towards its quantum-geometrical reformulation. 
Covariance would require us to treat even the time variables 
as  noncommutative objects and also to include 
all brane-antibrane dynamics automatically.  It would 
hopefully lay  the mathematical foundation for the space-time 
uncertainty relation. 

\vspace{0.3cm}
The present work grew out of discussions between us, 
begun when T.Y. was 
visiting Brown University under the US-Japan Collaborative Program 
for Scientific Research supported by the Japan Society for the  Promotion of 
Science.  He would like to thank Professor A. Jevicki for warm hospitality 
and discussions during his stay.  We would like to thank M. Douglas and
E. Martinec for reading the manuscript.
The work of M.L. 
was supported by DOE grant DE-FG02-90ER-40560 and NSF grant
PHY 91-23780.

\end{document}